\documentclass[preprintnumbers,superscriptaddress,showkeys]{revtex4}
\usepackage{amsmath,amsfonts,amssymb,amscd,amsxtra,amsthm}
\usepackage{graphicx}
\usepackage{epstopdf}
\usepackage{bm}
\usepackage{feynmp}
\usepackage{feynmp-auto}
\usepackage{orcidlink}
\usepackage{stackengine}
\begin{document}
\preprint{PKNU-NuHaTh-2026}
\title{Thermal modification of $K_1(1270)\to \pi^+\pi^-K^+$ in a hot hadronic medium}
\author{Seung-il Nam\,\orcidlink{0000-0001-9603-9775}}
\email[E-mail: ]{sinam@pknu.ac.kr}
\affiliation{Department of Physics, Pukyong National University (PKNU), Busan 48513, Korea}
\affiliation{Asia Pacific Center for Theoretical Physics (APCTP), Pohang 37673, Korea}
\date{\today}
\begin{abstract}
We study the thermal modification of the exclusive decay $K_1^+(1270)\to \pi^+\pi^-K^+$ in a hot hadronic medium. The decay amplitude is constructed from effective hadronic interactions dominated by the $\rho$- and $K^*$-pole contributions, which enables a Dalitz-level analysis of the three-body decay in medium. Thermal effects associated with partial chiral-symmetry restoration are incorporated through a phenomenological interpolation toward vector--axial-vector degeneracy near the chiral crossover. As the temperature increases, the reduction of the parent $K_1$ mass strongly compresses the available three-body phase space, leading to substantial deformation of the Dalitz distribution and invariant-mass spectra, as well as to a pronounced suppression of $\Gamma_{K_1\to\pi^+\pi^-K^+}(T)$. As a further step toward future experimental comparison, we introduce normalized shape observables that quantify the thermal evolution of the $K^*$-dominated $\pi K$ region, the upper-edge weight of the $\pi\pi$ spectrum, and the compactification of the Dalitz population. The dominant effect identified in the present framework is therefore kinematic: thermal phase-space reduction in the strange axial-vector channel. These results suggest that the exclusive $K_1(1270)$ channel may provide a useful qualitative probe of in-medium strange axial-vector dynamics near the pseudocritical region.
\end{abstract}
\maketitle
\section{Introduction}
Understanding how hadronic properties are modified in hot and dense QCD matter is a central goal of contemporary heavy-ion physics. Short-lived resonances are especially useful in this context because their observable yields and line shapes can be altered by in-medium decay, rescattering of decay products, and regeneration before kinetic freeze-out~\cite{Rapp:1999ej,ALICE:2017ban,ALICE:2021ptz}. Resonance measurements, therefore, provide sensitive probes of the late hadronic stage of relativistic nucleus-nucleus collisions and of the interplay between hadronization and subsequent hadronic evolution.

Experimentally, this sensitivity is illustrated by the suppression of short-lived resonances relative to longer-lived hadrons in central heavy-ion collisions. A representative example is the reduction of the $K^{*}(892)^0/K$ ratio with respect to the proton-proton baseline, commonly interpreted as evidence that hadronic rescattering reduces the reconstructable resonance signal~\cite{ALICE:2017ban,ALICE:2021ptz}. Similar suppression patterns observed for other short-lived states, such as $\Lambda(1520)$, further support the view that resonance yields reflect the lifetime and dynamical properties of the hadronic stage between chemical and kinetic freeze-out~\cite{Mosel:2010mp,Markert:2002rw,Rafelski:2001hp,ALICE:2018ewo}.

From the theoretical side, medium modifications of hadrons are closely related to the restoration of chiral symmetry. Since spontaneous chiral-symmetry breaking is a fundamental nonperturbative feature of low-energy QCD, its partial restoration at finite temperature is expected to induce substantial changes in hadronic spectral properties. The vector and axial-vector channels are of particular interest because chiral restoration implies that the corresponding chiral partners should tend toward degeneracy near the pseudocritical temperature~\cite{Rapp:1999ej,Tripolt:2021jtp,Weinberg:1967kj}. This idea has been widely explored in studies of spectral functions, dilepton production, and in-medium resonance properties, where the thermal evolution of vector and axial-vector mesons serves as an important indicator of changing chiral dynamics~\cite{Rapp:1999ej}. In this context, the enhancement of the $K_1/K^*$ ratio in heavy-ion collisions has also been proposed as a possible signature of hot-medium effects and chiral dynamics~\cite{Sung:2023oks,Sung:2021myr}.

Against this background, the strange axial-vector meson $K_1(1270)$ is particularly interesting. As a strange member of the axial-vector sector, it provides a complementary perspective on in-medium chiral dynamics beyond the nonstrange channels. Its dominant hadronic decays proceed through vector-meson intermediate states, making it sensitive both to the thermal evolution of the strange sector and to the available hadronic phase space. At the same time, the physical $K_1(1270)$ state is closely related to the well-known $K_1(1270)$-$K_1(1400)$ mixing problem, which has been studied in a variety of phenomenological and QCD-based approaches~\cite{Cheng:2013cwa,Shi:2023kiy}. These features make the $K_1(1270)$ meson a potentially useful, though still relatively unexplored, probe of hot hadronic matter.

In this work, we investigate the three-body decay process $K_1^+(1270)\to \pi^+\pi^-K^+$ in a hot hadronic medium. We first construct the decay amplitude in vacuum within an effective hadronic framework that includes the dominant $\rho$- and $K^*$-pole contributions. We then introduce thermal modifications of the relevant vector and axial-vector mesons through a phenomenological parametrization inspired by the Weinberg sum rules, such that the chiral-partner pairs $(\rho,a_1)$ and $(K^*,K_1)$ gradually approach degeneracy as the temperature increases. Within this framework, we examine how the Dalitz structure, invariant-mass distributions, and the total three-body decay width are modified in a medium.

Since chemical freeze-out occurs near the QCD crossover region, medium-induced modifications of the parent axial-vector state may leave characteristic traces in exclusive hadronic decay channels before the system reaches kinetic freeze-out~\cite{Andronic:2017pug,Chatterjee:2015fua}. The temperature dependence of the $K_1$ decay width is therefore relevant for understanding how the strange axial-vector channel responds to the hot hadronic environment created in relativistic nucleus-nucleus collisions.

Most previous discussions have focused either on inclusive resonance-yield ratios, such as $K_1/K^*$, or on global in-medium properties of vector and axial-vector mesons, rather than on an exclusive three-body channel with full Dalitz-level kinematics. In contrast, we treat the decay $K_1(1270)^+ \to \pi^+\pi^-K^+$ itself as a temperature-dependent observable. Our goal is to isolate the leading kinematic consequence of partial chiral restoration for a reconstructable strange axial-vector decay channel by combining an explicit hadronic decay amplitude with a phenomenological description of in-medium vector--axial-vector evolution. In addition, we introduce normalized shape observables constructed from the daughter invariant-mass spectra and the Dalitz distribution itself, to provide experimentally oriented measures of thermal distortion that are less sensitive to overall normalization uncertainty.

The present framework is intentionally exploratory. It is not designed to provide a quantitative description of experimentally reconstructed $K_1$ yields in heavy-ion collisions, since regeneration, daughter-particle rescattering, coupled-channel dynamics, detector acceptance, and combinatorial backgrounds are not included. Instead, the aim is to identify robust qualitative signatures of thermal phase-space suppression in the exclusive channel $K_1(1270)^+ \to \pi^+\pi^-K^+$. This paper is organized as follows: In Sec.~II, we formulate the decay in vacuum and introduce the temperature-dependent meson properties. In addition, the shape observables are also defined. In Sec.~III, we present numerical results for Dalitz plots, invariant-mass spectra, thermal decay width, etc., with detailed discussions. Section~IV contains the summary and outlook.

\section{Theoretical Framework}

\subsection{$K_1^+\to\pi^+\pi^-K^+$ decay}
\begin{figure}[t]
\includegraphics[width=13cm]{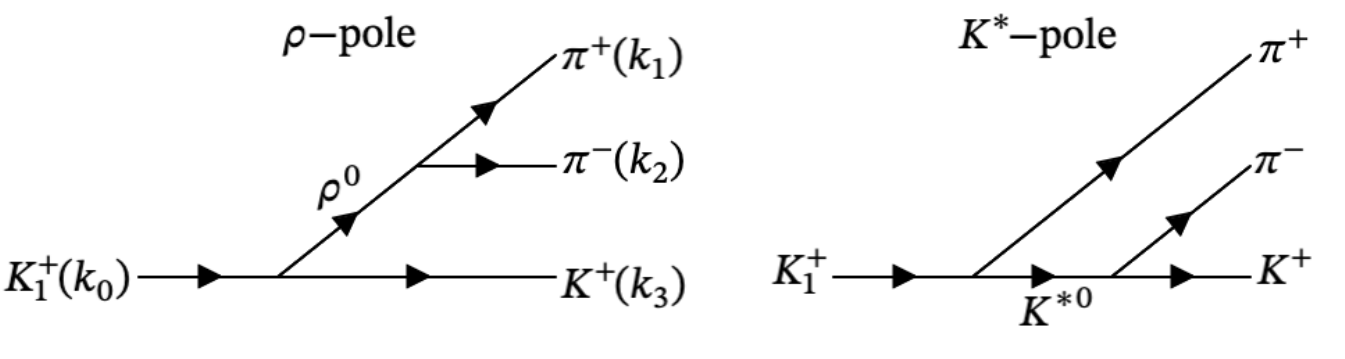}
\caption{Relevant Feynman diagrams for the decay $K_1^+\to \pi^+\pi^-K^+$, namely, the $\rho$- and $K^*$-pole contributions. The four-momentum assignments are also indicated.}
\label{FIG0}
\end{figure}

In this subsection, we formulate the vacuum decay $K_1^+(1270)\to \pi^+\pi^-K^+$, which serves as the baseline for the finite-temperature analysis below. We assume that the dominant contributions arise from the two resonant processes shown in Fig.~\ref{FIG0}, namely the $\rho$-pole and $K^*$-pole channels. These correspond to the sequential decays $K_1^+\to \rho K^+$ followed by $\rho\to\pi^+\pi^-$, and $K_1^+\to K^*\pi$ followed by $K^*\to K\pi$, respectively. The decay amplitude is described within an effective hadronic framework with axial-vector, vector, and pseudoscalar meson degrees of freedom.

The relevant interaction vertices are introduced through the effective Lagrangians
\begin{eqnarray}
\label{eq:EFL}
\mathcal{L}_{VPP}=ig_{VPP}V^\mu\left(P\partial_\mu P^\dagger-P^\dagger\partial_\mu P\right)+\mathrm{h.c.},
\qquad
\mathcal{L}_{AVP}=ig_{AVP}A^\mu V_\mu P+\mathrm{h.c.},
\end{eqnarray}
where $A$, $V$, and $P$ denote the axial-vector, vector, and pseudoscalar meson fields, respectively. The interaction $\mathcal{L}_{VPP}$ generates the $\rho\pi\pi$ and $K^*K\pi$ vertices, while $\mathcal{L}_{AVP}$ describes the $K_1\rho K$ and $K_1K^*\pi$ transitions. Because of the derivative structure of $\mathcal{L}_{VPP}$, the resulting amplitudes depend on momentum differences of the outgoing pseudoscalar mesons, as expected for intermediate spin-1 states.

Using Eq.~(\ref{eq:EFL}), one obtains the invariant amplitudes for the $\rho$- and $K^*$-pole diagrams in Fig.~\ref{FIG0} as
\begin{eqnarray}
\label{eq:AMPS}
i\mathcal{M}_{\rho}
=\frac{g_{K\rho K_1}g_{\pi\pi\rho}\left(\epsilon\cdot q_{1-2}\right)}{q^2_{1+2}-M^2_\rho+i\Gamma_\rho M_\rho},
\qquad
i\mathcal{M}_{K^*}=\frac{g_{\pi K^*K_1}g_{\pi KK^*}\left[\epsilon\cdot q_{2-3}-\left(\frac{M^2_\pi-M^2_K}{M^2_{K^*}}\right)\epsilon\cdot q_{2+3}\right]}{q^2_{2+3}-M^2_{K^*}+i\Gamma_{K^*} M_{K^*}},
\end{eqnarray}
where $q_{i\pm j}\equiv k_i\pm k_j$, with $k_1$, $k_2$, and $k_3$ denoting the four-momenta of $\pi^+$, $\pi^-$, and $K^+$, respectively. The intermediate $\rho$ meson carries $q_{1+2}^2=(k_1+k_2)^2=M_{\pi^+\pi^-}^2$, whereas the intermediate $K^*$ meson carries $q_{2+3}^2=(k_2+k_3)^2=M_{\pi^-K^+}^2$. The denominators are of Breit-Wigner form and encode the finite widths of the unstable intermediate resonances.

The momentum structure in Eq.~(\ref{eq:AMPS}) follows directly from the derivative $VPP$ interaction. In the $\rho$-pole channel, the $\rho\to\pi^+\pi^-$ vertex yields $(k_1-k_2)^\mu\equiv q_{1-2}^\mu$, so that the numerator is proportional to $\epsilon\cdot q_{1-2}$. In the $K^*$-pole channel, the transverse part of the vector propagator gives the expected term $\epsilon\cdot q_{2-3}$, while the longitudinal projector $q_\mu q_\nu/M_{K^*}^2$ with $q=q_{2+3}$ generates the additional contribution proportional to $\epsilon\cdot q_{2+3}$. Using
\begin{equation}
q\cdot (k_2-k_3)=(k_2+k_3)\cdot(k_2-k_3)=M_\pi^2-M_K^2,
\end{equation}
one obtains the mass-dependent correction appearing in the numerator of $\mathcal{M}_{K^*}$. Numerically, this term is subleading, but it is retained for completeness and preserves the covariant form of the amplitude in the unequal-mass $K\pi$ channel.

\begin{table*}[b]
\begin{tabular}{c|cc|cccc}
\hline\hline
State 
& $g_{\pi K^* K_1}$~[GeV] 
& $g_{K\rho K_1}$~[GeV]
& $g_{\pi K K^*}$ 
& $g_{\pi\pi\rho}$ 
& $\Gamma_{K^*}$~[MeV] 
& $\Gamma_{\rho}$~[MeV] \\
\hline
$K_1(1270)$ & 0.92 & 4.10 & 4.48 & 8.49 & 47.3 & 147.7 \\
$K_1(1400)$ & 4.34 & 0.92 & \multicolumn{4}{c}{same as above} \\
\hline\hline
\end{tabular}
\caption{Relevant coupling constants and widths for $K_1(1270)$ and $K_1(1400)$ decays.}
\label{TAB1}
\end{table*}

The total decay amplitude is written as the coherent sum
\begin{eqnarray}
\label{eq:fullamp}
\mathcal{M}_{\rho+K^*}(\epsilon)=\mathcal{M}_{\rho}(\epsilon)+\mathcal{M}_{K^*}(\epsilon)+\mathcal{M}_{\rm BKG},
\end{eqnarray}
so that the interference between the two resonant channels is fully included. Here, $\mathcal{M}_{\mathrm{BKG}}$ denotes a constant nonresonant background term introduced to absorb smooth contributions not explicitly described by the dominant $\rho$- and $K^*$-pole amplitudes. In the present phenomenological treatment, this term is not meant to represent a specific microscopic mechanism, but rather to parametrize residual direct three-body or broad nonresonant effects at the amplitude level. Its value will be fixed to reproduce a reasonable $K_1$ partial decay width in Sec.~III. For completeness, the propagator of an intermediate vector meson with four-momentum $q$ is given by
\begin{eqnarray}
\label{eq:vecprop}
iD_{V\mu\nu}(q)=\frac{i}{q^2-M_V^2+i\Gamma_V M_V}\left(-g_{\mu\nu}+\frac{q_\mu q_\nu}{M_V^2}\right).
\end{eqnarray}

For the initial $K_1$ meson at rest, we choose the polarization vectors as~\cite{Park:2024vga}
\begin{eqnarray}
\label{eq:polvec}
\epsilon_+=\left(0,-\frac{1}{\sqrt{2}},-\frac{i}{\sqrt{2}},0\right),
\qquad
\epsilon_0=(0,0,0,1),
\qquad
\epsilon_-=\left(0,\frac{1}{\sqrt{2}},-\frac{i}{\sqrt{2}},0\right).
\end{eqnarray}
These satisfy the standard orthonormality and completeness relations for a massive spin-1 particle. In the decay-width calculation, the squared amplitude is averaged over the three spin states of the initial $K_1$, which leads to the factor $1/3$ in the expression below.

The strong coupling constants are taken from Refs.~\cite{Sung:2021myr,ParticleDataGroup:2024cfk}, where the physical $K_1(1270)$ and $K_1(1400)$ states were analyzed, including their mixing structure. The numerical values used here are listed in Table~\ref{TAB1}. Since these couplings already encode the phenomenological effects of $K_1(1270)$-$K_1(1400)$ mixing, we do not introduce additional mixing parameters in the present work. Our purpose is instead to examine how thermal phase-space reduction modifies the decay once a phenomenologically acceptable set of physical couplings is adopted.

To describe the three-body kinematics, we use the two independent invariant masses $M_{\pi^+\pi^-}^2=(k_1+k_2)^2$ and $M_{\pi^-K^+}^2=(k_2+k_3)^2$, which span the Dalitz plot. The double-differential decay width is then given by
\begin{eqnarray}
\label{eq:DDG}
\Gamma_{K_1(1270)\to \pi^+\pi^-K^+}\equiv\frac{d^2\Gamma_{K_1(1270)\to \pi^+\pi^-K^+}}
{dM_{\pi^+\pi^-}dM_{\pi^-K^+}}
=\frac{1}{3}\sum_{x=0,\pm1}\frac{M_{\pi^+\pi^-}M_{\pi^-K^+}}{64\pi^3M^3_{K_1}}
|\mathcal{M}_{\rho+K^*+{\rm BKG}}|^2.
\end{eqnarray}
This expression contains the full resonance and interference structure of the three-body decay in terms of the two Dalitz variables. The factor $M_{\pi^+\pi^-}M_{\pi^-K^+}/(64\pi^3M_{K_1}^3)$ arises from the Lorentz-invariant three-body phase-space measure written in terms of the two independent invariant masses. Integration of Eq.~(\ref{eq:DDG}) over the kinematically allowed Dalitz region yields the total vacuum decay width.

This vacuum formalism provides the starting point for the finite-temperature analysis. Once thermal modifications of hadron masses and widths are incorporated, the pole positions and available phase space in Eq.~(\ref{eq:AMPS}) are altered accordingly, which leads to corresponding changes in both the Dalitz distribution and the total decay width.

\subsection{Temperature-dependent meson properties}
To model thermal modifications of the vector ($V$) and axial-vector ($A$) mesons, we adopt a phenomenological framework motivated by partial chiral restoration. Our qualitative guidance comes from the first and second Weinberg sum rules,
\begin{eqnarray}
\label{eq:WSR}
\int_{0}^{\infty} ds \left[\rho_V(s)-\rho_A(s)\right]=f_\pi^2,
\qquad
\int_{0}^{\infty} ds\, s \left[\rho_V(s)-\rho_A(s)\right]=0,
\end{eqnarray}
where $\rho_V(s)$ and $\rho_A(s)$ denote the vector and axial-vector spectral functions, respectively, and $f_\pi$ is the pion weak-decay constant. As the system approaches the chiral-restoration region, the difference between the vector and axial-vector channels is expected to diminish. In the present work, however, these sum rules are used only as qualitative motivation for the tendency toward vector--axial-vector degeneracy, i.e., we do not attempt a microscopic calculation of in-medium spectral functions. Our thermal setup is therefore constructed as an effective interpolation scheme designed to isolate the leading kinematic consequence of axial-vector mass reduction for the exclusive three-body decay $K_1^+\to\pi^+\pi^-K^+$. The purpose is not to provide a precise description of strange-sector spectroscopy in medium, but to examine how a smooth crossover-driven approach toward vector--axial-vector degeneracy modifies the available Dalitz phase space and the resulting decay width.

We adopt a simplified description in which the pseudoscalar meson masses are kept approximately constant, $M_P(T)\approx M_P(0)$ with $P=\pi,K$, while the vector-meson masses and widths are also assumed to remain close to their vacuum values, $M_V(T)\approx M_V(0)$ and $\Gamma_V(T)\approx \Gamma_V(0)$. This choice reflects the expectation that, in the temperature range relevant to the present study, the dominant thermal effect arises from the axial-vector sector, whereas the pseudoscalar mesons are comparatively less sensitive to the medium. Although the kaon is not an exact Nambu-Goldstone boson and vector mesons may acquire collision-induced broadening in a realistic hadronic environment, these effects are expected to be subleading for the present purpose, namely to isolate the leading phase-space suppression induced by the reduction of the parent axial-vector mass. Mesonic thermal properties in the PNJL model were studied in Ref.~\cite{Carlomagno:2019yvi}, which exhibit a trend qualitatively consistent with the assumptions adopted here: the axial-vector channel changes more strongly than the vector channel.

To parametrize the temperature dependence of the axial-vector masses, we assume that their evolution follows the thermal crossover profile of the constituent light-quark mass, $M_A(T)\propto M_{q=u,d}(T)$, where $M_q(T)$ denotes the effective light-quark mass at temperature $T$. To determine $M_q(T)$, we employ the two-flavor Nambu--Jona-Lasinio (NJL) model in medium. The NJL Lagrangian is given by
\begin{eqnarray}
\label{eq:NJL}
\mathcal{L}_{\mathrm{NJL}}=\bar{q}(i\gamma^\mu\partial_\mu-m_q)q
+G\left[(\bar{q}q)^2+(\bar{q}i\gamma_5\bm{\tau}q)^2\right],
\end{eqnarray}
where $q=(u,d)^T$ denotes the light-quark doublet, $m_q$ the current quark mass, and $G$ the effective $2N_f$ coupling constant. Although the NJL model does not incorporate confinement, it captures the essential mechanism of spontaneous chiral-symmetry breaking through dynamical mass generation. The constituent quark mass is determined self-consistently as
\begin{eqnarray}
\label{eq:MQQ}
M_q=m_q-2G\langle\bar{q}q\rangle.
\end{eqnarray}

At finite temperature $T$ and quark chemical potential $\mu_q$, the constituent mass is obtained in the Matsubara formalism as
\begin{align}
\label{eq:mqTmu}
M_q(T,\mu_q)=m_q-2G\langle\bar{q}q\rangle,\,\,\,
\langle\bar{q}q\rangle=-4N_cN_f\int_0^\Lambda\frac{d^3\vec{k}}{(2\pi)^3}\frac{M_q}{E_k}\left[1-n_+-n_-\right],
\end{align}
where
\begin{eqnarray}
E_k=\sqrt{\vec{k}^{\,2}+M_q^2},\,\,\,
n_F(E_k\pm\mu_q)=\frac{1}{e^{(E_k\pm\mu_q)/T}+1}\equiv n_\mp.
\end{eqnarray}
As $T$ or $\mu_q$ increases, the quark condensate decreases, which reduces $M_q$ and signals partial chiral restoration.

Here, we focus on the $\mu_q=0$ case and use the standard NJL parameter set $m_q=5.25~\mathrm{MeV}$, $\Lambda=631.4~\mathrm{MeV}$, and $G\Lambda^2=2.14$~\cite{Klevansky:1992qe}. We have also checked the parameter dependence by varying $\Lambda=(630\pm30)$ MeV and $G\Lambda^2=2.1\pm0.1$, and find that the qualitative behavior and the pseudocritical region remain essentially unchanged. In the present analysis, the two-flavor NJL input is used solely to provide a smooth, normalized crossover profile of $M_q(T)$. The strange-sector mass scale itself is not predicted by the model, but is introduced phenomenologically through an interpolation between $M_A$ and its vector partner $M_V$. The corresponding interpretation of this construction in the strange axial-vector sector is discussed in the Appendix.

To implement this idea in a compact form, we parametrize the normalized quark-mass profile $M_q(T)/M_q(0)$ by a Richards-type generalized logistic function,
\begin{eqnarray}
\label{eq:Richards}
F(T)=A+B\left[1+\exp\left(\frac{T-T_{0}}{\Delta T}\right)\right]^{-\nu},
\end{eqnarray}
where $A=0.08299$, $B=0.9180$, $\Delta T=0.02668\,\mathrm{GeV}$, and $\nu=0.1988$. Here, $T_0$ [GeV] is a fit parameter controlling the midpoint of the crossover profile. Using this function, the axial-vector mass is interpolated between its vacuum value and the corresponding vector-meson mass as
\begin{eqnarray}
\label{eq:MAT}
M_A(T)=M_V+\left[M_A(0)-M_V\right]
\left[\frac{F(T)-F(T_{\mathrm{pc}})}{F(0)-F(T_{\mathrm{pc}})}\right].
\end{eqnarray}
Here, $T_{\mathrm{pc}}$ denotes the pseudocritical temperature that sets the onset of approximate vector--axial-vector degeneracy in the present model. Throughout this work, we use $T_c$ and $T_{\mathrm{pc}}$ interchangeably as the same crossover scale, with $T_c \approx  T_{\mathrm{pc}} \approx  1.84~\mathrm{GeV}$. This construction ensures that $M_A(T)$ approaches $M_V$ as the system moves toward the chiral-restoration region. We have also verified that alternative smooth parametrizations, such as hyperbolic-tangent forms, do not qualitatively change the numerical results, indicating that the dominant conclusion is controlled by the existence of a smooth crossover profile rather than by the particular functional form adopted.

The thermal decay widths are then updated consistently with the changing kinematics. Rather than keeping the axial-vector widths fixed, we assume that their dominant temperature dependence originates from the available two-body phase space for the decay $A\to VP$. We therefore write
\begin{eqnarray}
\label{eq:GA_T}
\Gamma_A(T)=\Gamma_A(0)
\left[\frac{M_A(0)}{M_A(T)}\right]
\left[\frac{p_A(T)}{p_A(0)}\right]^3,
\qquad
\Gamma_V(T)\approx \Gamma_V(0),
\end{eqnarray}
where the factor $p_A^3(T)$ reflects the $P$-wave nature of the decay. The center-of-mass momentum is given by
\begin{eqnarray}
\label{eq:pA_T}
p_A(T)=\frac{\sqrt{\left[M_A^2(T)-(M_V+M_P)^2\right]\left[M_A^2(T)-(M_V-M_P)^2\right]}}{2M_A(T)}.
\end{eqnarray}
As $M_A(T)$ decreases toward $M_V$, the available phase space shrinks, and the axial-vector width is correspondingly reduced.

The approximation $\Gamma_V(T)\approx \Gamma_V(0)$ also deserves comment. In a realistic hadronic medium, vector mesons such as the $\rho$ and $K^*$ are expected to acquire additional collision-induced broadening. However, the dominant effect in the present analysis comes from the reduction of the parent axial-vector mass $M_{K_1}(T)$, which shrinks the available Dalitz domain and suppresses the $A\to VP$ phase space entering Eq.~(\ref{eq:GA_T}). By contrast, moderate thermal broadening of the intermediate $K^*$ mainly redistributes spectral strength around the resonance pole and smooths the corresponding ridge structure in the Dalitz plot, without reversing the overall reduction of the total three-body phase space caused by the decreasing $K_1$ mass. Neglecting vector broadening is therefore not expected to alter the qualitative trend, although it can affect detailed line shapes and should be incorporated in a more quantitative treatment. Nonetheless, we will test a mild broadening of vector-meson widths in the numerical calculations, as shown in Sec.~III. 

A similar remark applies to the approximation $M_P(T)\approx M_P(0)$. Even if moderate thermal shifts of the pseudoscalar masses are allowed, their impact on the present decay kinematics is expected to remain subleading compared with the axial-vector mass reduction. The three-body threshold for $K_1^+\to \pi^+\pi^-K^+$ is $2M_\pi+M_K\approx 0.77~\mathrm{GeV}$ in vacuum, whereas the decrease of $M_{K_1}(T)$ in Fig.~\ref{FIG1}(a) is of order $0.2~\mathrm{GeV}$ over the temperature range considered. By comparison, even a conservative thermal increase of $(10$--$20)~\mathrm{MeV}$ per pseudoscalar would raise the threshold only by $(30$--$60)~\mathrm{MeV}$. Keeping the pseudoscalar masses fixed is therefore a reasonable first approximation for isolating the dominant phase-space effect, while explicit thermal pseudoscalar-mass shifts are left for future refinement.

Within this framework, the thermal modification of the $K_1$ mass and width directly affects the three-body decay $K_1^+\to\pi^+\pi^-K^+$ through the intermediate $\rho$- and $K^*$-pole channels. Both the resonance structure and the kinematically allowed Dalitz region are therefore altered at finite temperature. As we show below, the resulting decay width $\Gamma_{K_1\to\pi^+\pi^-K^+}(T)$ is strongly suppressed with increasing temperature, primarily because the decreasing parent axial-vector mass rapidly contracts the available three-body phase space.

\subsection{Experiment-oriented shape observables}
While the present framework is not intended to provide quantitative predictions for experimentally reconstructed heavy-ion yields, it is nevertheless useful to define normalized observables that facilitate future comparisons with exclusive reconstructed data. This is particularly important because the present analysis isolates decay-level thermal distortions more reliably than absolute yield modifications. Since the dominant thermal effect identified here is the shrinking of the available Dalitz support, the most natural observables are shape-based quantities that are less sensitive to the overall normalization uncertainty associated with the nonresonant background.

A first example is the fraction of events inside a $K^*$-dominated window of the $M_{\pi^-K^+}$ spectrum,
\begin{eqnarray}
R_{K^*}(T;\Delta_{K^*})
&\equiv&
\frac{\int_{M_{K^*}-\Delta_{K^*}}^{M_{K^*}+\Delta_{K^*}}
dM_{\pi^-K^+}\frac{d\Gamma(T)}{dM_{\pi^-K^+}}}{\int dM_{\pi^-K^+}\frac{d\Gamma(T)}{dM_{\pi^-K^+}}}.
\label{eq:RKS}
\end{eqnarray}
This quantity measures how strongly the reconstructed event population remains concentrated around the resonance-enhanced $K^*$ region as the temperature increases. A second useful observable is the relative spectral weight near the upper edge of the $M_{\pi^+\pi^-}$ distribution,
\begin{eqnarray}
R_{\mathrm{edge}}(T;M_{\mathrm{cut}})
&\equiv&
\frac{\int_{M_{\mathrm{cut}}}^{M_{\pi\pi}^{\max}(T)}
dM_{\pi^+\pi^-}\frac{d\Gamma(T)}{dM_{\pi^+\pi^-}}}{\int dM_{\pi^+\pi^-}\frac{d\Gamma(T)}{dM_{\pi^+\pi^-}}}.
\label{eq:RKE}
\end{eqnarray}
This observable is designed to capture the progressive suppression of the upper-bound enhancement in the $\pi^+\pi^-$ channel as the available phase space becomes compressed.

If full Dalitz-level information is available, one may also define a compactification measure of the event population through an intensity-weighted second moment,
\begin{eqnarray}
C(T)&\equiv&\frac{\int dM_{\pi^+\pi^-}\, dM_{\pi^-K^+}W(M_{\pi^+\pi^-},M_{\pi^-K^+})
\frac{d^2\Gamma(T)}{dM_{\pi^+\pi^-}dM_{\pi^-K^+}}}{\int dM_{\pi^+\pi^-}\, dM_{\pi^-K^+}
\frac{d^2\Gamma(T)}{dM_{\pi^+\pi^-}dM_{\pi^-K^+}}},
\label{eq:CD}
\end{eqnarray}
where a natural choice for the weight function is
\begin{eqnarray}
W(M_{\pi^+\pi^-},M_{\pi^-K^+})=[M_{\pi^+\pi^-}-\bar M_{\pi\pi}^{(0)}]^2
+[M_{\pi^-K^+}-\bar M_{\pi K}^{(0)}]^2,
\end{eqnarray}
with $\bar M_{\pi\pi}^{(0)}$ and $\bar M_{\pi K}^{(0)}$ denoting the centroid of the vacuum Dalitz distribution. In this form, $C(T)$ quantifies the thermal contraction of the event population relative to the vacuum kinematic support.

For practical implementation, we adopt the reference choices
\begin{eqnarray}
\Delta_{K^*}=50~\mathrm{MeV},
\qquad
M_{\mathrm{cut}}=0.65~\mathrm{GeV},
\end{eqnarray}
which are intended to capture, respectively, the resonance-dominated region around the $K^*$ peak in the $M_{\pi^-K^+}$ spectrum and the upper-edge enhancement region in the $M_{\pi^+\pi^-}$ spectrum.  These observables are not meant to replace a full transport-based description of experimentally reconstructed heavy-ion signals. Rather, they provide intermediate decay-level quantities that are directly sensitive to the weakening and narrowing of the resonance-enhanced structures and to the shrinking of the Dalitz support. In this sense, they offer a practical bridge between the present phenomenological analysis and future comparisons with reconstructed exclusive data.

\section{Numerical Results}
In this section, we present the numerical results obtained within the temperature-dependent framework described above and discuss their implications for the decay process $K_1^+(1270)\to\pi^+\pi^-K^+$ in a hot hadronic medium. Our focus is on the thermal deformation of the Dalitz structure, the invariant-mass spectra, the integrated three-body decay width, and the normalized shape observables.

Before turning to the thermal results, we estimate the background amplitude $\mathcal{M}_{\rm BKG}$. Since the exclusive partial width for the channel $K_1(1270)^+\to K^+\pi^+\pi^-$ is not listed directly in the PDG~\cite{ParticleDataGroup:2024cfk}, we infer a representative value from the total width and the dominant quasi-two-body branching fractions. As a bookkeeping estimate, we decompose the charged final state into the leading intermediate channels,
\begin{eqnarray}
\mathcal{B}_{\rm est}[K_1(1270)^+ \to K^+\pi^+\pi^-]
&\approx &
\mathcal{B}[K_1^+ \to K^+\rho^0]\mathcal{B}[\rho^0\to\pi^+\pi^-]
\cr
&+& 
\mathcal{B}[K_1^+ \to K^{*0}\pi^+]\mathcal{B}[K^{*0}\to K^+\pi^-]
\cr
&+& 
\mathcal{B}[K_1^+ \to K_0^{*0}(1430)\pi^+]\mathcal{B}[K_0^{*0}(1430)\to K^+\pi^-],
\end{eqnarray}
where only the charge combinations contributing to the final state $K^+\pi^+\pi^-$ are retained. The corresponding partial width is approximated as
\begin{equation}
\Gamma_{\rm est}^{\rm PDG}\!\left[K_1(1270)^+ \to K^+\pi^+\pi^-\right]
=
\mathcal{B}_{\rm est}\!\left[K_1(1270)^+ \to K^+\pi^+\pi^-\right]
\Gamma_{K_1(1270)}^{\rm tot}.
\end{equation}
Using $\Gamma_{K_1(1270)}^{\rm tot}\approx 101~\mathrm{MeV}$ together with the PDG-listed branching fractions for the dominant quasi-two-body channels, we obtain an estimated branching fraction $\mathcal{B}_{\rm est}\left[K_1(1270)^+ \to K^+\pi^+\pi^-\right]\approx 0.4$, within the corresponding experimental uncertainties. Accordingly,
\begin{equation}
\Gamma_{\rm est}^{\rm PDG}\!\left[K_1(1270)^+ \to K^+\pi^+\pi^-\right]\approx 40~\mathrm{MeV}.
\end{equation}

\begin{figure}[t]
\topinset{(a)}{\includegraphics[width=7.5cm]{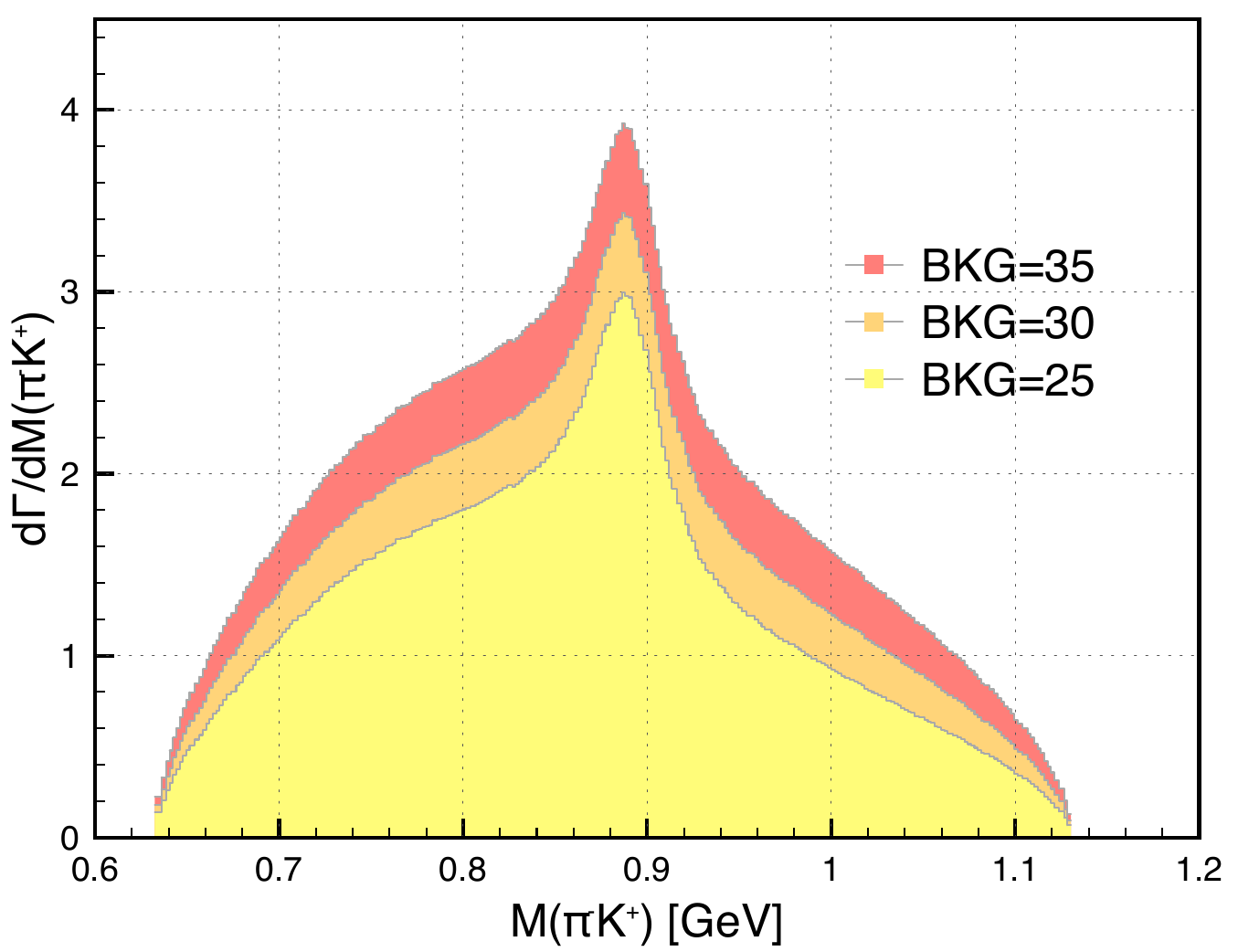}}{-0.4cm}{0.0cm}
\caption{Invariant-mass distribution $d\Gamma/dM(\pi^-K^+)$ for $K_1(1270)^+\to\pi^+\pi^-K^+$ with $\mathcal{M}_{\rm BKG}=25$, 30, and 35 at $T=0$. The background mainly changes the overall strength, while the spectral shape remains nearly unchanged.}
\label{FIGBKG}
\end{figure}

This estimate should be regarded as only indicative, since it neglects coherent interference among the intermediate resonances as well as possible nonresonant contributions. In the full amplitude-level treatment of Eq.~(\ref{eq:AMPS}), by contrast, the $\rho$- and $K^*$-pole terms interfere coherently over the entire Dalitz domain. Numerically, we obtain $\Gamma\left[K_1(1270)^+\to \pi^+\pi^-K^+\right]\approx (32$--$47)~\mathrm{MeV}$ for $\mathcal{M}_{\rm BKG}\approx (25$--$35)$. In the following analysis, we adopt $\mathcal{M}_{\rm BKG}=30$, which reproduces the PDG-based estimate at a reasonable level: $\Gamma\left[K_1(1270)^+\to \pi^+\pi^-K^+\right]\approx39.5$ MeV.  We have also checked that varying the background strength within this range does not produce any significant change in the Dalitz-plot topology. As shown in Fig.~\ref{FIGBKG}, the $\pi^-K^+$ invariant-mass distribution for $\mathcal{M}_{\rm BKG}=25$, 30, and 35 at $T=0$ exhibits mainly an overall enhancement of the spectral strength, while its resonance structure remains essentially unchanged.

The absolute normalization, therefore, retains some background dependence, but the decreasing trend with temperature remains stable within this range. Although a more complete treatment of the nonresonant amplitude may modify the detailed normalization and local Dalitz-profile intensity, the qualitative temperature dependence obtained here is not driven by the specific choice of $\mathcal{M}_{\rm BKG}$ but by the kinematic reduction of the available three-body phase space as $M_{K_1}(T)$ decreases.

\begin{figure}[t]
\topinset{(a)}{\includegraphics[width=7.5cm]{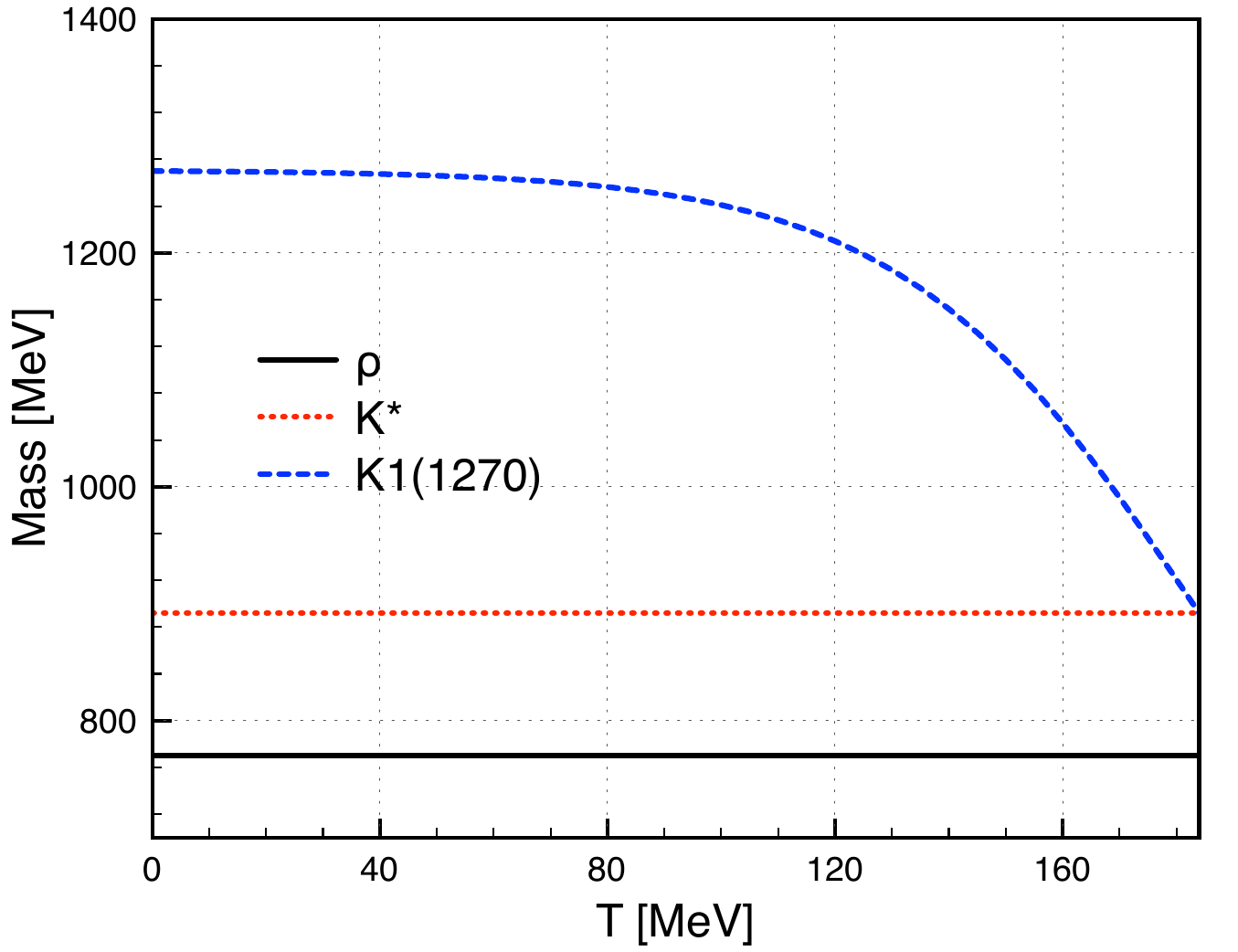}}{-0.4cm}{0.0cm}
\topinset{(b)}{\includegraphics[width=7.5cm]{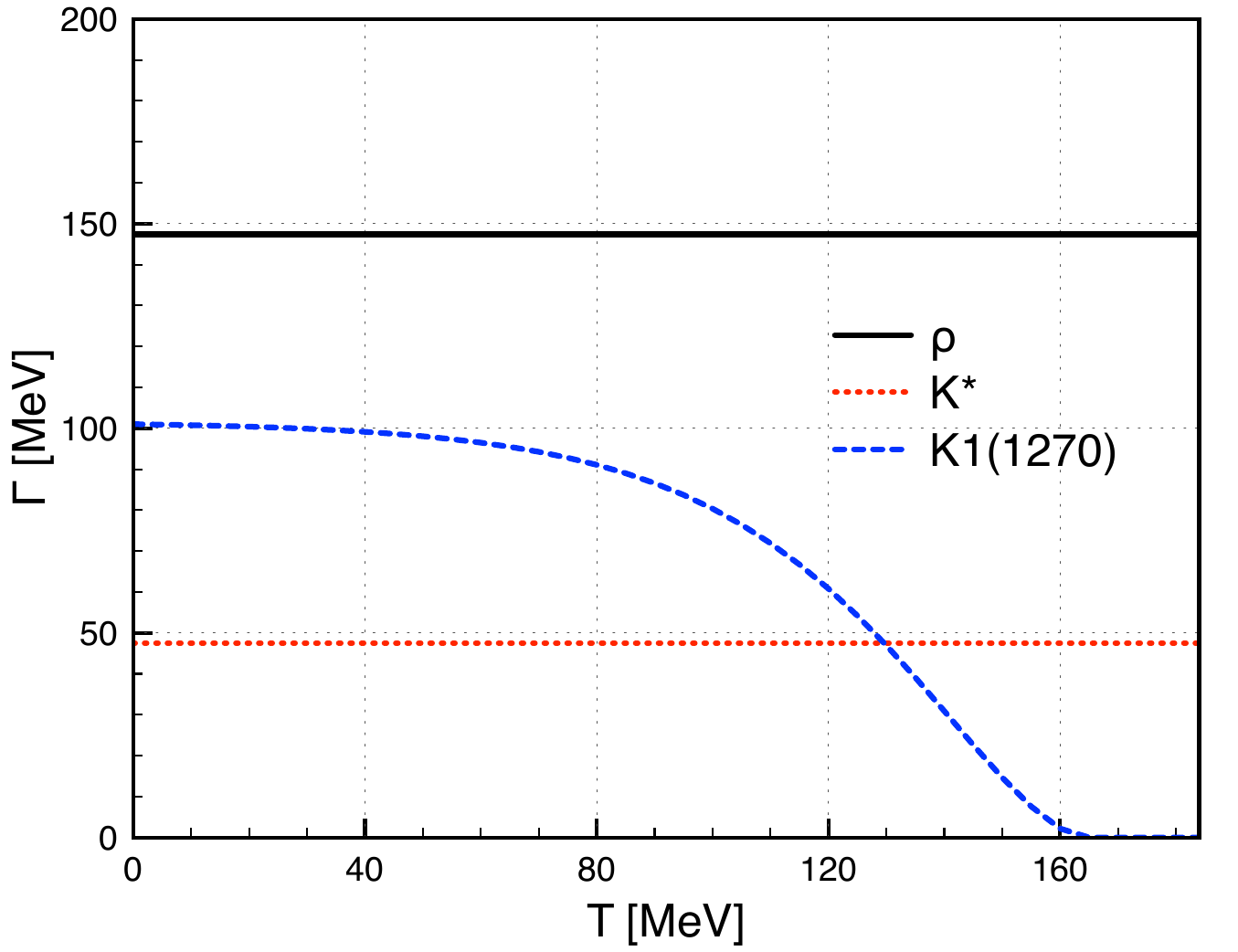}}{-0.4cm}{0.0cm}
\caption{Temperature dependence of the meson properties used in the present calculation: (a) masses of the $\rho$, $K^*$, and $K_1(1270)$ mesons, and (b) their corresponding decay widths. The more rapid variation in the range $T =(120$--$160)~\mathrm{MeV}$ reflects the approach to the pseudocritical region, where approximate vector--axial-vector degeneracy sets in, and the $K_1$ channel begins to show strong suppression.}
\label{FIG1}
\end{figure}

We first examine the thermal behavior of the meson properties entering the present calculation. Fig.~\ref{FIG1} shows the temperature dependence of the masses and decay widths of the relevant vector and axial-vector mesons. By construction, the vector-meson masses remain nearly constant, whereas the axial-vector masses decrease with temperature and gradually approach their corresponding vector partners. The associated axial-vector widths are also strongly reduced because they are updated through the shrinking phase space for the $A\to VP$ channel. In the present parametrization, this evolution remains mild at low temperature and becomes much steeper only in the pseudocritical region. For the $K_1$ channel in particular, the reduction of both $M_{K_1}(T)$ and $\Gamma_{K_1}(T)$ becomes pronounced around $T=(120$--$130)~\mathrm{MeV}$ and accelerates toward chemical freeze-out.

\begin{figure}[t]
\begin{tabular}{ccc}
\topinset{(a) $T=0$ MeV}{\includegraphics[width=5.5cm]{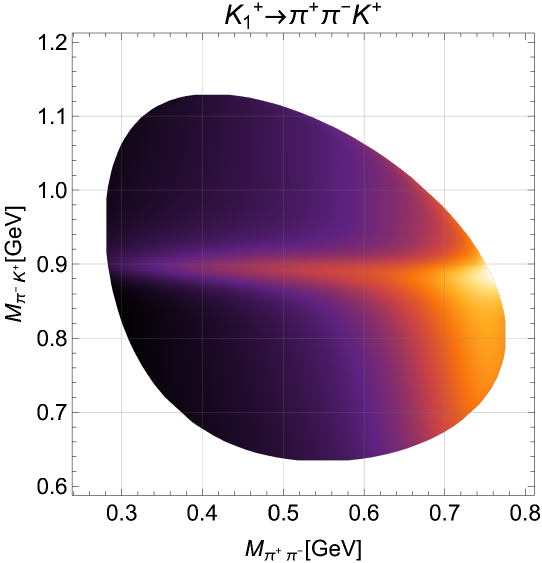}}{-0.4cm}{0.0cm}
\topinset{(b) $T=110$ MeV}{\includegraphics[width=5.5cm]{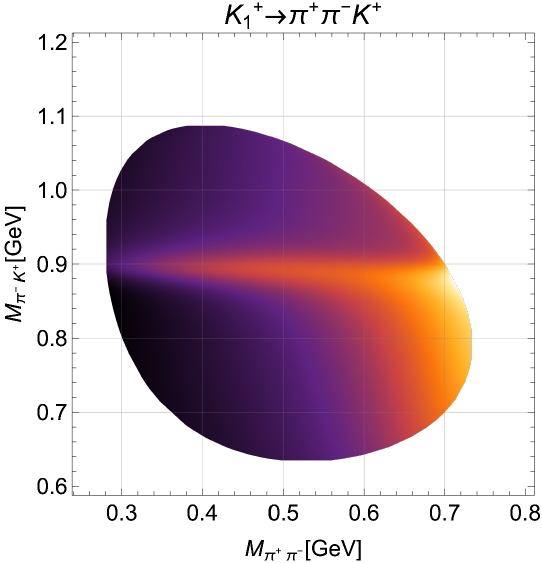}}{-0.4cm}{0.0cm}
\topinset{(c) $T=130$ MeV}{\includegraphics[width=5.5cm]{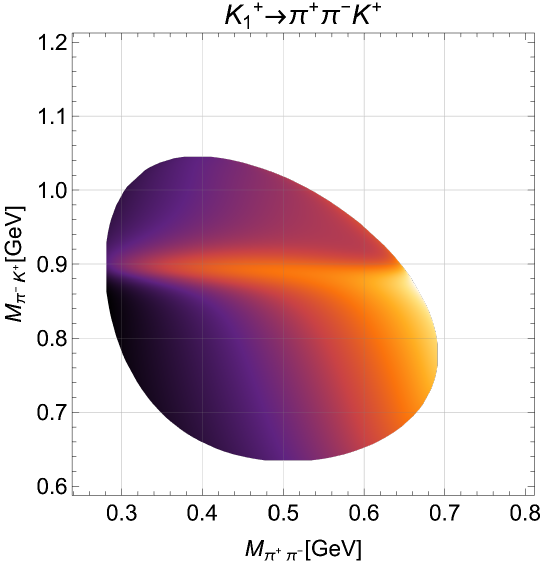}}{-0.4cm}{0.0cm}
\\
\topinset{(d) $T=140$ MeV}{\includegraphics[width=5.5cm]{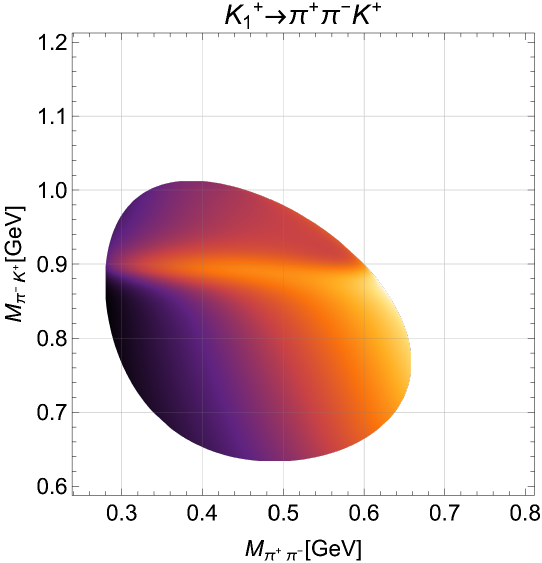}}{-0.4cm}{0.0cm}
\topinset{(e) $T=150$ MeV}{\includegraphics[width=5.5cm]{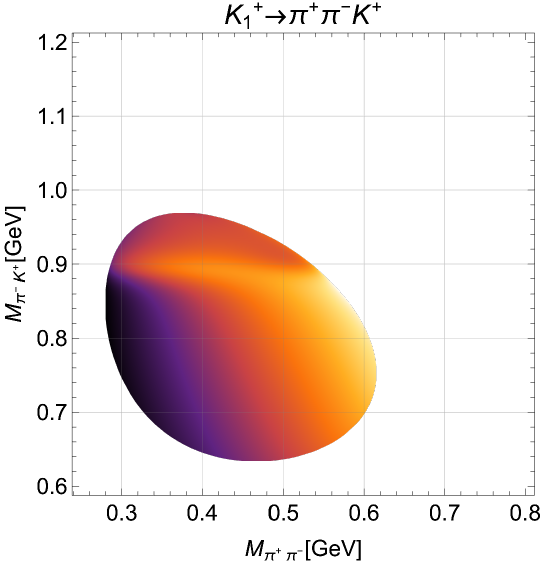}}{-0.4cm}{0.0cm}
\topinset{(f) $T=160$ MeV}{\includegraphics[width=5.5cm]{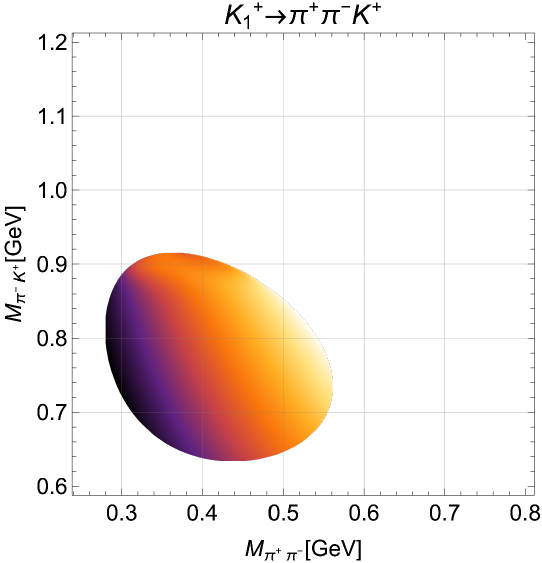}}{-0.4cm}{0.0cm}
\end{tabular}
\caption{Dalitz plots for $K_1^+(1270)\to \pi^+\pi^-K^+$ as functions of the invariant masses $M_{\pi^+\pi^-}$ and $M_{\pi^-K^+}$, shown with a common color normalization for $T=(0$--$160$) MeV.}
\label{FIG2}
\end{figure}

The modification of the meson properties is reflected directly in the Dalitz distribution for the three-body decay $K_1^+\to\pi^+\pi^-K^+$. Fig.~\ref{FIG2} shows the Dalitz plots for $T=(0$--$160$) MeV. At zero temperature, the allowed kinematic region is broad and the event distribution exhibits a pronounced ridge associated mainly with the $K^*$-pole contribution, together with a weaker structure related to the $\rho$ channel. As the temperature increases, the Dalitz domain shrinks steadily because the decreasing $K_1$ mass reduces the total energy available to the final-state particles. At the same time, the high-intensity ridge becomes narrower and shifts toward smaller invariant masses. The medium, therefore, modifies both the overall decay strength and the detailed kinematic profile through shifted pole positions and reduced phase space.

\begin{figure}[t]
\topinset{(a)}{\includegraphics[width=7.5cm]{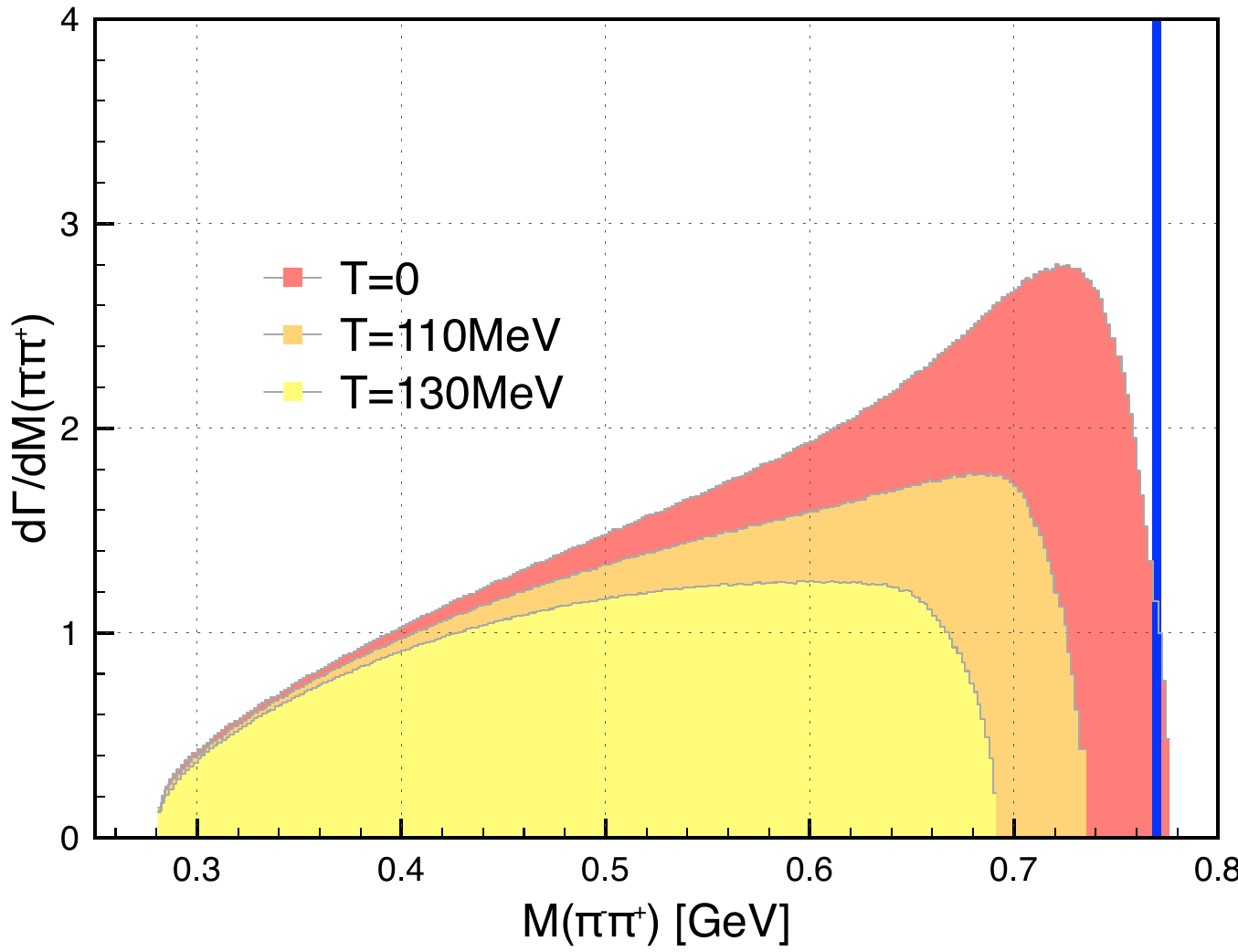}}{-0.4cm}{0.0cm}
\topinset{(b)}{\includegraphics[width=7.5cm]{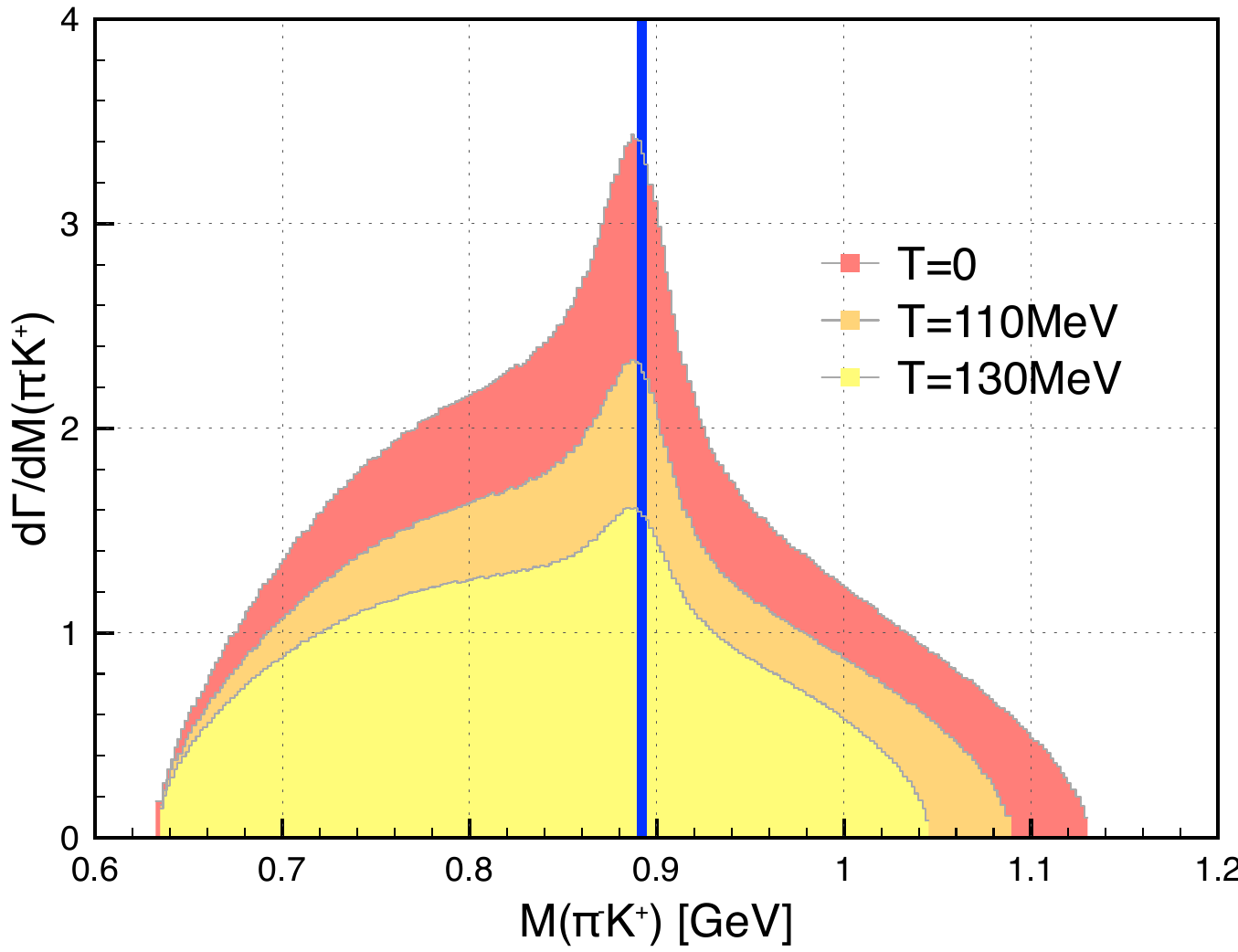}}{-0.4cm}{0.0cm}
\caption{Invariant-mass distributions for $K_1^+\to \pi^+\pi^-K^+$ at several temperatures: (a) $d\Gamma/dM_{\pi^+\pi^-}$ and (b) $d\Gamma/dM_{\pi^-K^+}$. The vertical thick lines stand for the mass of $\rho^0(770)$ and $K^{*0}(892)$.}
\label{FIG3}
\end{figure}

The same trend appears in the one-dimensional invariant-mass distributions shown in Fig.~\ref{FIG3}. In the $M_{\pi^+\pi^-}$ distribution (a), the enhancement near the upper kinematic boundary is progressively suppressed as $T$ increases. Likewise, the $M_{\pi^-K^+}$ spectrum (b) shows a pronounced peak around the $K^*$ region at low temperature, but this peak becomes reduced and compressed at higher temperature. These results indicate that thermal modification affects not only the integrated strength of the decay but also the shapes of experimentally relevant invariant-mass spectra.

\begin{figure}[t]
\begin{tabular}{ccc}
\topinset{(a)}{\includegraphics[width=5.5cm]{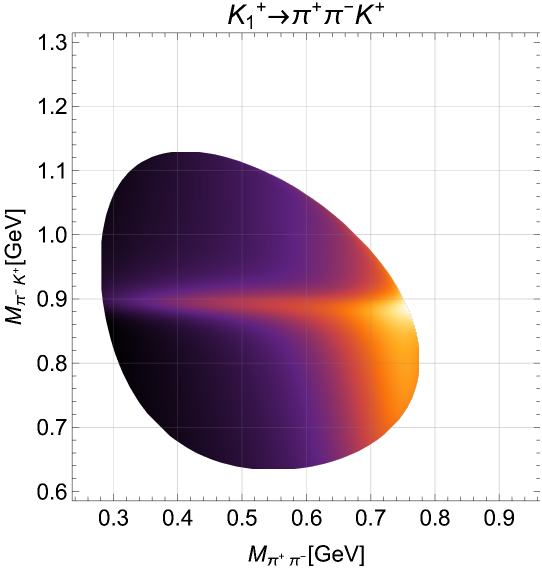}}{-0.4cm}{0.0cm}
\topinset{(b)}{\includegraphics[width=5.5cm]{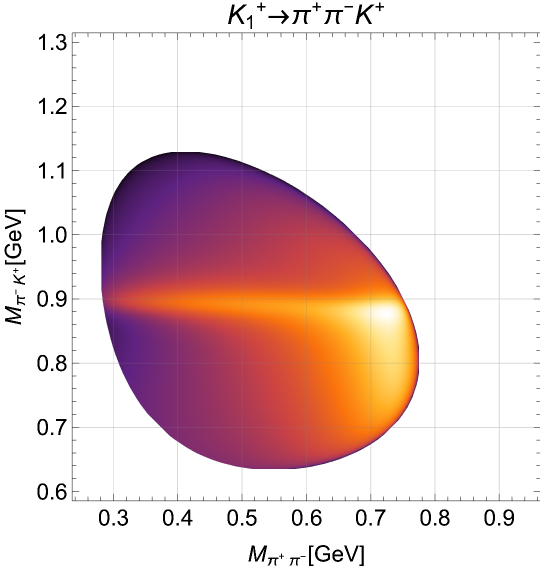}}{-0.4cm}{0.0cm}
\\
\topinset{(c)}{\includegraphics[width=5.5cm]{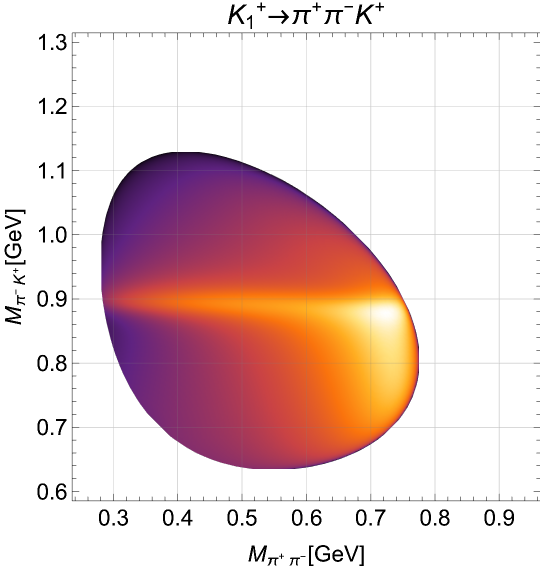}}{-0.4cm}{0.0cm}
\topinset{(d)}{\includegraphics[width=5.5cm]{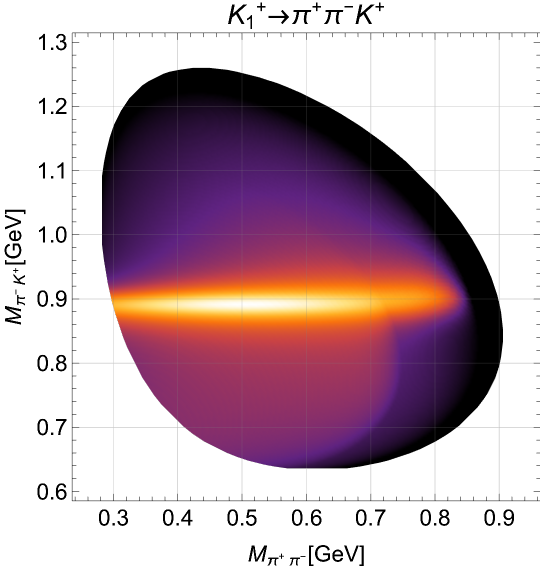}}{-0.4cm}{0.0cm}
\end{tabular}
\caption{(a) Dalitz plot at $T=0$, (b) thermally averaged one for $T=(0$--$T_c)$, and (c) that with the vector-meson width broadening in Eq.~(\ref{eq:BRO}). (d) Thermally averaged one including $K_1(1400)$.}
\label{FIG4}
\end{figure}

To illustrate the cumulative effect of the thermal evolution, Fig.~\ref{FIG4}(a) and (b) compare the vacuum Dalitz plot with the thermally averaged one, respectively, over the range $T=(0$--$T_c)$. The thermally averaged distribution is visibly smeared and reduced relative to the vacuum case, reflecting the fact that decays occurring at different temperatures sample different effective masses, widths, and phase-space boundaries. If a mild broadening of the vector-meson width is taken into account as indicated in the PNJL model calculation~\cite{Carlomagno:2019yvi}, the widths for $\rho^0$ and $K^{*0}$ can be parameterized simply as follows:
\begin{eqnarray}
\Gamma_V(T)=\Gamma_V\left[1+0.2\left(\frac{T}{T_c}\right)^2\right].
\label{eq:BRO}
\end{eqnarray}

Fig.~\ref{FIG4}(c) shows the thermally-averaged Dalitz plot with the thermal broadening of $\Gamma_V$ as in Eq.~(\ref{eq:BRO}), and it turns out that the mild broadening make negligible difference in comparison to that without it. Fig.~\ref{FIG4}(d) provides a purely illustrative cross-check obtained by including the $K_1(1400)$ contribution. Although the detailed ridge intensity and overall event distribution are modified by admixture with the heavier strange axial-vector state, the main thermal trend remains unchanged: the resonance-enhanced structure is smeared, and the suppression associated with the shrinking effective phase space persists. In the present paper, the $K_1(1400)$ contribution is included only for illustration, rather than as part of a systematic in-medium treatment of $K_1(1270)$-$K_1(1400)$ mixing.

\begin{figure}[t]
\topinset{(a)}{\includegraphics[width=7.5cm]{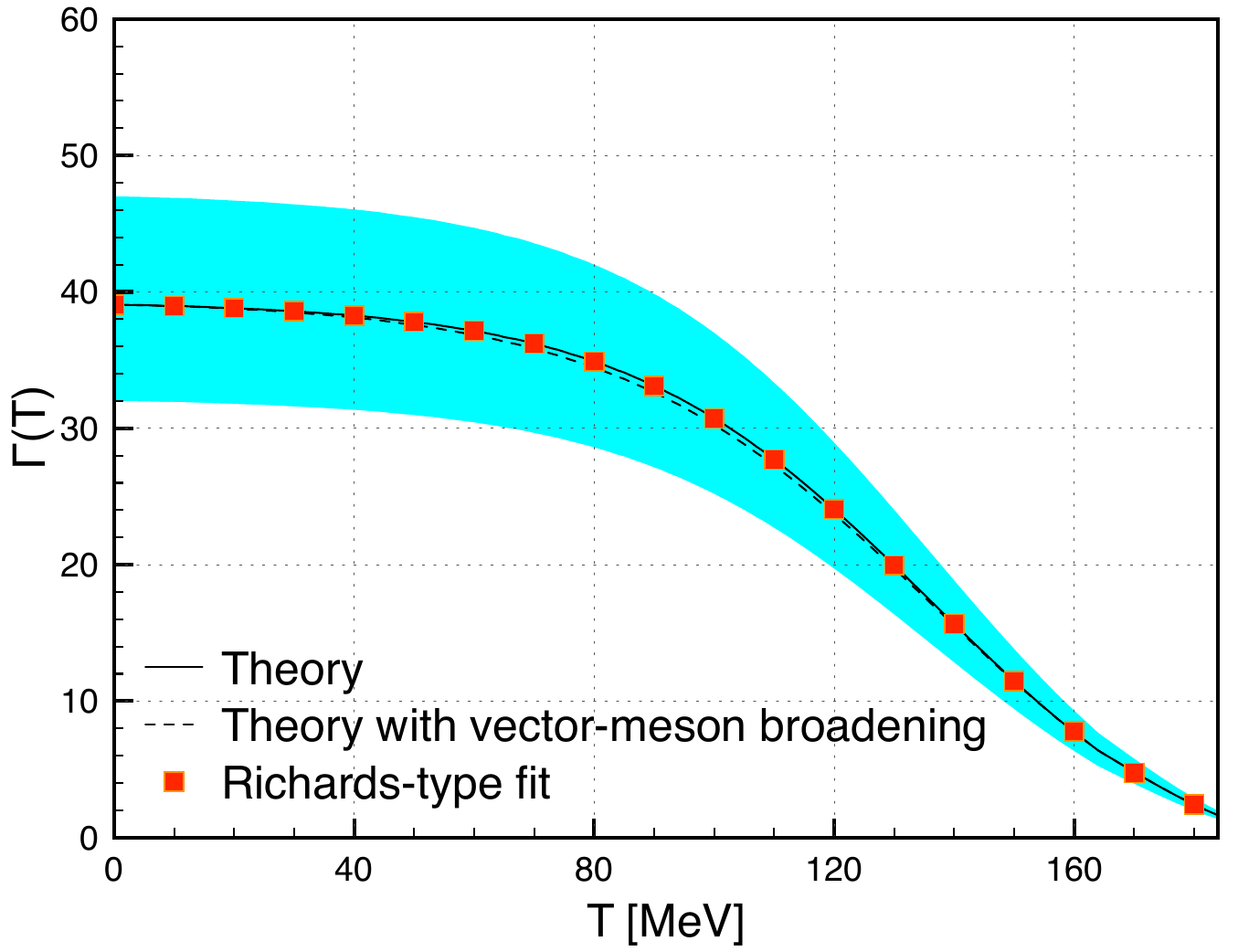}}{-0.4cm}{0.0cm}
\topinset{(b)}{\includegraphics[width=7.5cm]{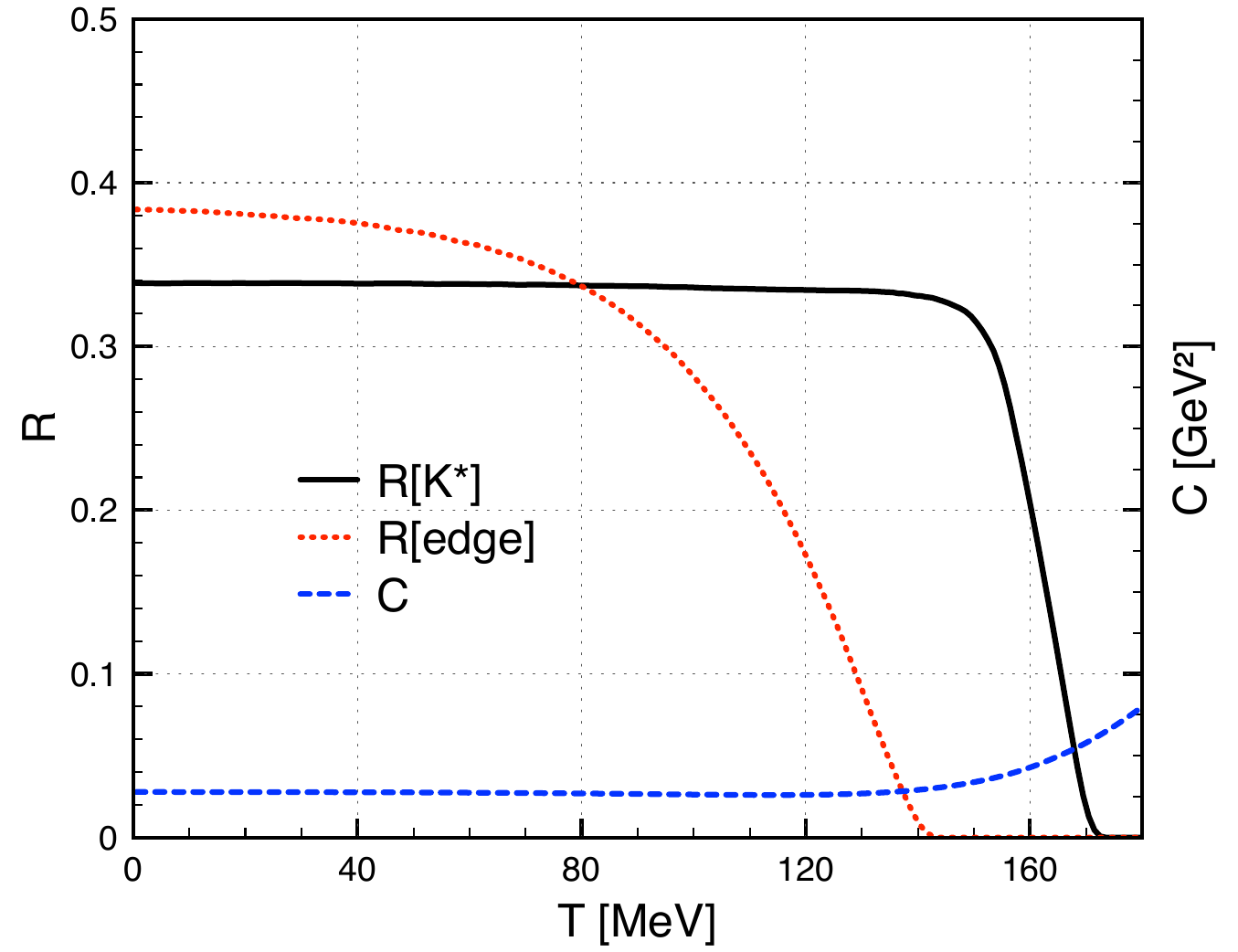}}{-0.4cm}{0.0cm}
\caption{(Color online) (a) Temperature dependence of the integrated decay width $\Gamma_{K_1\to\pi^+\pi^-K^+}(T)$ for the central choice $\mathcal{M}_{\rm BKG}=30$, together with the Richards-type fit. The shaded band indicates the variation obtained for $\mathcal{M}_{\rm BKG}=(25$--$35)$. (b) Normalized shape observables $R_{K^*}$, $R_{\mathrm{edge}}$, and $C$, defined by Eqs.~(\ref{eq:RKS}-\ref{eq:CD}).}
\label{FIG5}
\end{figure}

Integrating the double-differential decay width over the Dalitz region yields the total three-body decay width $\Gamma_{K_1\to\pi^+\pi^-K^+}(T)$ for the central choice $\mathcal{M}_{\rm BKG}=30$, shown in Fig.~\ref{FIG5}(a). The solid and dashed lines denote the decay widths with and without the vector-meson width broadening, respectively, defined by Eq.~(\ref{eq:BRO}). As expected from Fig.~\ref{FIG4}(c), the broadening does not make any obvious differences. The shaded band indicates the variation obtained by changing the nonresonant background amplitude in the range $\mathcal{M}_{\rm BKG}=(25$--$35)$. The result is well described by the Richards-type parametrization (square):
\begin{eqnarray}
\label{eq:GammaK1_fit}
\Gamma_{K_1^+\to \pi^+\pi^- K^+}(T)=\Gamma_{K_1^+\to \pi^+\pi^- K^+}(0)
\left[a+b\left[1+\exp\left(\frac{T-T_c}{\Delta T}\right)\right]^{-\nu}\right],
\end{eqnarray}
with $a=-2.461$, $b=41.73$, $T_c=142.4\,\mathrm{MeV}$, $\Delta T=25.90\,\mathrm{MeV}$, and $\nu=1.292$. The suppression becomes increasingly pronounced toward the pseudocritical region, where the reduction of $M_{K_1}(T)$ rapidly compresses the available three-body phase space. The dominant conclusion is therefore kinematic: within the present phenomenological framework, the thermal suppression of the exclusive channel is controlled primarily by the shrinking Dalitz volume induced by the decreasing axial-vector mass. 

To complement this integrated measure, we also evaluate the normalized shape observables introduced in Eqs.~(\ref{eq:RKS}-\ref{eq:CD}). These quantities allow one to track how the thermal suppression is distributed across different regions of the daughter invariant-mass spectra and the Dalitz plane, rather than only through the total decay strength. Fig.~\ref{FIG5}(b) summarizes the thermal evolution of the normalized shape observables. The upper-edge weight $R_{\mathrm{edge}}$ decreases earlier and more rapidly than the $K^*$-window fraction $R_{K^*}$, indicating that the enhancement near the upper boundary of the $M_{\pi^+\pi^-}$ spectrum is more sensitive to the initial compression of the available phase space. By contrast, the resonance-dominated $K^*$ structure in the $M_{\pi^-K^+}$ channel remains relatively stable up to higher temperature and is strongly reduced only near the pseudocritical region. Meanwhile, the increase of $C$ shows that the Dalitz event population is progressively redistributed as the allowed kinematic support shrinks and deforms. These results demonstrate that the thermal effect identified here manifests itself not only as an overall suppression of the decay width but also as a nontrivial reshaping of normalized exclusive observables. 

Overall, the numerical results show that thermal reduction of the available Dalitz phase space deforms the invariant-mass spectra and significantly suppresses the integrated decay width of $K_1^+\to\pi^+\pi^-K^+$. This indicates that the $K_1(1270)$ channel can serve as a qualitative probe of in-medium strange axial-vector dynamics and the approach to chiral restoration. From an experimental perspective, the present results imply not only an overall suppression of reconstructable $K_1(1270)^+ \to \pi^+\pi^-K^+$ candidates, but also characteristic modifications of the daughter invariant-mass spectra. As the parent axial-vector mass decreases, the three-body phase space becomes increasingly compressed, driving events into a narrower kinematic region. In particular, the $M_{\pi^-K^+}$ distribution is expected to show a weakened and compressed $K^*$-dominated structure, while the enhancement in the $M_{\pi^+\pi^-}$ spectrum near the upper kinematic boundary becomes progressively less pronounced. The thermal effect is therefore reflected not only in the integrated width, but also in qualitative shape changes that may remain visible in reconstructed exclusive channels.

\section{Summary and outlook}
We have investigated the three-body decay process $K_1^+(1270)\to\pi^+\pi^-K^+$ in a hot hadronic medium. The decay was formulated in vacuum within an effective hadronic framework that includes the dominant $\rho$- and $K^*$-pole contributions, providing a basis for analyzing the Dalitz structure, invariant-mass spectra, and the integrated three-body decay width. Thermal effects were incorporated through a phenomenological framework motivated by partial chiral-symmetry restoration, in which vector--axial-vector chiral partners gradually approach degeneracy toward the pseudocritical region. In this setup, reducing the $K_1$ mass leads to a substantial contraction of the available three-body phase space and, consequently, a strong suppression of the decay width.

Our numerical analysis shows that both the Dalitz plot and the corresponding one-dimensional invariant-mass spectra are significantly modified with increasing temperature. The kinematically allowed region becomes progressively smaller, while the resonance-enhanced structures, particularly those associated with the $K^*$ channel, become narrower and weaker. The medium, therefore, affects not only the overall magnitude of the decay rate but also the detailed kinematic profile of the decay. To characterize this effect beyond the integrated width, we introduced normalized shape observables that quantify the thermal evolution of the $K^*$-dominated $\pi K$ region, the upper-edge weight of the $\pi\pi$ spectrum, and the compactification of the Dalitz population. These observables demonstrate that the thermal modification identified here is not exhausted by an overall suppression of the decay rate, but also appears as a structured reshaping of the exclusive decay kinematics.

The thermal setup adopted here is intentionally phenomenological and is not derived from a microscopic finite-temperature calculation of strange spectral functions. Its purpose is to isolate the leading decay-level consequence of a smooth reduction of the $K_1$-$K^*$ mass splitting as the system approaches vector--axial-vector degeneracy. Under this interpretation, we expect the sign and overall trend of the suppression to be more robust than its detailed magnitude, whereas the quantitative size of the effect and the detailed line shapes may still depend on thermal broadening in the intermediate vector channels, finite-temperature pseudoscalar-mass shifts, and a more microscopic treatment of the strange-sector thermal input.

Several limitations of the present analysis should be noted. First, the nonresonant background term $\mathcal{M}_{\mathrm{BKG}}$ was introduced only phenomenologically as a temperature-independent contribution and should not be regarded as a microscopic description of direct three-body dynamics. Although a more realistic treatment could affect quantitative details of the Dalitz distribution and the partial width, the present results suggest that the dominant thermal effect is driven primarily by the reduction in the available three-body phase space as $M_{K_1}(T)$ decreases.

Second, the present treatment of the $K_{1}(1270)$-$K_{1}(1400)$ system remains schematic. While the adopted vacuum couplings already contain phenomenological information related to their mixing, a dynamical in-medium evolution of the mixing pattern has not been included~\cite{Suzuki:1993yc}. A more complete treatment of the strange axial-vector sector would therefore be desirable in future work.

Even within this simplified framework, however, the present results already point to qualitative experimental signatures at the level of reconstructed exclusive candidates. In addition to an overall reduction in the decay strength, the shrinking Dalitz support implies characteristic distortions in the invariant-mass distributions. These shape changes may provide useful observables that are less sensitive to overall normalization uncertainties. A more quantitative connection to experimentally reconstructed heavy-ion observables will require implementing the in-medium spectral modifications in realistic dynamical transport simulations, such as UrQMD~\cite{Bleicher:1999xi} or SMASH~\cite{SMASH:2016zqf}. Related works are in progress and will appear elsewhere. 

\section*{Acknowledgment}
The author is grateful to Su-Jeong Ji, Sanghoon Lim, and Su Houng Lee (Yonsei University) for fruitful discussions. This work was supported by the National Research Foundation of Korea (NRF) grant funded by the Korean government (MSIT) (RS-2025-16065906).

\section*{Appendix}
\section{Phenomenological implementation of the $N_f=2$ NJL crossover profile\\ in the strange axial-vector sector}
In the present work, the two-flavor NJL model is not used as a microscopic theory for the in-medium strange-meson spectrum. It is employed only as a source of a smooth, normalized crossover profile through $M_q(T)/M_q(0)$. The strange-sector mass scale itself is not predicted by the model, but is introduced phenomenologically through
\begin{equation}
M_A(T)=M_V+\bigl[M_A(0)-M_V\bigr]
\left[
\frac{F(T)-F(T_{\mathrm{pc}})}{F(0)-F(T_{\mathrm{pc}})}
\right].
\end{equation}
Thus, the NJL input fixes only the location and smoothness of the crossover, not the absolute strange-meson spectroscopy.

The motivation is that the $K_1$ and $K^*$ states contain one light quark, so their vector--axial-vector splitting is expected to respond qualitatively to light-sector chiral restoration in the same pseudocritical region. The present ansatz should therefore be understood as a phenomenological crossover-scale mapping rather than a microscopic strange-sector model. Its purpose is to isolate the dominant kinematic consequence of decreasing $K_1$-$K^*$ splitting. In the exclusive channel studied here, the main effect is the rapid contraction of the available three-body phase space as $M_{K_1}(T)$ decreases, which suppresses the decay width. A more microscopic treatment may change the detailed thermal trajectory, but the overall suppression trend is expected to remain as long as the effective $K_1$-$K^*$ splitting decreases smoothly toward the pseudocritical region, because the dominant effect is the associated phase-space contraction.

\end{document}